\documentclass[twocolumn,showpacs,showkeys,preprintnumbers,amsmath,amssymb,prb,floatfix]{revtex4}

\usepackage{graphicx}
\usepackage{dcolumn}
\usepackage{bm}
\usepackage{psfrag}

\begin{document}


\title{Quench dynamics in spin crossover induced by high pressure}

\author{A. I. Nesterov}
   \email{nesterov@cencar.udg.mx}
\affiliation{Departamento de F{\'\i}sica, CUCEI, Universidad de Guadalajara,
Av. Revoluci\'on 1500, Guadalajara, CP 44420, Jalisco, M\'exico}

\author{S. G. Ovchinnikov}
 \email{sgo@iph.krasn.ru}
\affiliation{L. V. Kirensky Institute of Physics, SB RAS, 660036, Krasnoyarsk , Russia \\
Institute of Engeniering Physics and Radioelectronics, Siberian Federal University, Krasnoyarsk 660041, Russia }

\date{\today}

\begin{abstract}
In this paper we have studied analytically and numerically dynamics of spin crossover induced by time-dependent pressure. We show that quasi static pressure, with a slow dependence on time, yields a spin crossover leading to transition from the state of quantum system with high spin (HS) to the low spin (LS). However, a quench dynamics under shock-wave load is more complicated. The final state of the system depends on the amplitude and pulse velocity, resulting in the mixture of
the HS and LS states. 
\end{abstract}



\maketitle

\section{Introduction}

Spin crossover in condensed matter physics is a transformation of a system with one spin $S_1$ at each lattice site into another state with spin $S_2$ induced by some external parameter like strong magnetic field, high pressure etc. It accompanies by the energy level $E_1$ and $E_2$ crossing, where $E_a$ is the local energy of the magnetic ion with spin $S_a$ ($a=1,2$). Recently spin crossovers in magnetic oxides have been found under high pressure in $FeBO_3$ \cite{STL},  $CdFe_3(BO_3)_4 $ \cite{GKL},  $BiFeO_3$  \cite{GSL},  $Fe_3O_4$ \cite{DHO}. Below the Curie temperature of magnetic order spin crossover is accompanied by the sharp change of the magnetization, nevertheless it may be observed in the paramagnetic state like in $CdFe_3(BO_3)_4$ \cite{GKL} as the sharp change of the XES satellite/main peak intensity ratio with pressure increase. 

Energy levels crossing results in the lost of analyticity in the energy spectrum at the critical point (in the thermodynamic limit) \cite{SS}. Near the critical point adiabaticity breaks down and non-equilibrium phenomena associated with the drastically grown quantum fluctuations can drive the system away from the ground state. The final result depends on how fast the transition occurs. If the quench process is sufficiently fast, large numbers of topological defects are created and the final state, being characterized by mixture of high and low spin phases, can be essentially different from that been obtained as result of slow evolution. Qualitatively, quench dynamics can be described by the  Kibble-Zurek theory of nonequilibrium phase transitions \cite{KTW,ZHW,ZHW1}.

In this paper we consider quench dynamics in spin crossover induced by time-dependent pressure. The paper is organized as follows. In Sec. II, a general model of spin crossover under high pressure is introduced. In Sec. III, we study quench dynamics. We consider two cases: a) Pressure is a linear function of time; b) Pressure is defined by a pulse of a given shape. We conclude in Sec. IV with a discussion of our results.

\section{Model}

The multielectron ion in a crystal field has the energies of terms for $d^n$  configurations determined numerically by the Tanabe-Sugano diagrams \cite{TS} as a solution of the eigenvalue problem. Simple analytical calculations of the low energy terms with different spin value that is sufficient to study spin crossover has been done recently \cite{SGO}. The crystal field parameter increase linearly with pressure $P$. Thus the multielectron energies for spin  $S_1$  and  $S_2$  ( $E_1$  and  $E_2$ ) are also linear functions of $P$. To distinguish two different spin states in the lattice we introduce the Ising  pseudospin states $|i\rangle$  and  $|-i\rangle$   for  $|d_i^n,S^i_1\rangle$ and $|d_i^n,S^i_2\rangle$, where $i$ runs over all sites in the lattice. Thus we neglect the spin degeneracy of the $d_i^n$ terms but capture the possibility of energy level crossing that is the essential part of the spin crossover. Then, in the basis  $|+i\rangle$, $|-i\rangle$, the Hamiltonian of the system can be written as follows
\begin{align}\label{H1a}
H= \sum_i\big({\lambda^i_0}{1\hspace{-.125cm}1} +{\varepsilon_i}\hat\sigma^z_{i}) + \sum_{ij}H_{ij},
\end{align}
where $\lambda^i_0=(E^i_1+  E^i_2)/2$, $\varepsilon_i =(E^i_1 -  E^i_2)/2 $, and  ${1\hspace{-.125cm}1} $, $\hat\sigma_z$ are the identity and Pauli matrices, respectively; the Hamiltonian of interaction between the spins being $H_{ij}$.

The $H_{ij}$  includes the isotropic Heisenberg term with the exchange interaction $I_{ij}$ between nearest spins and the anisotropic term $H_A$. The interatomic interaction $I_{ij}$ is negligibly small in comparison with the interatomic Hund's coupling (ratio $~ 10^{-2}$). Thus its contribution to the localized spin energy $E_1$ and $E_2$  due to the effective molecular field can be neglected. Nevertheless the exchange interaction plays very important role: it results in the long range order and synchronize each spin in the same quantum state providing a cooperative behavior of the spin system. If it were the ferromagnetic interaction, each spin at T=0 would have the maximal projection $S$ with integer magnetic moment $2S$.

In all examples given above there is the antiferromagnetic interaction. The ground state of the isotropic Heisenberg antiferromagnet has non integer local magnetic moment due to the quantum spin fluctuations. It is known that for large spin $S$ effect of quantum fluctuations is less important then for small spin, and for $FeBO_3$ spin is $5/2$. Moreover the magnetic anisotropy additionally suppresses quantum fluctuations. For example, in $FeBO_3$ the anisotropy field is $0.3T$ \cite{VLV} and the measured value of the effective moment $2\sqrt{S(S+1)} = 5.9$  is very close to the calculated for $S=5/2$ value $5.916$.

Thus we conclude that due to anisotropy the magnetic moment at $T=0$ has integer value ( of course it is a property of the magnetic insulator that does not hold for itinerant magnets), and due to exchange interaction all spins are in the same quantum state. So spin crossover at $T=0$  is the transition of the whole crystal from one magnetically ordered state to another. Nevertheless the criterium of the transition can be found from consideration of the single ion energies crossover due to space uniform cooperative magnetic order.
Anisotropic relativistic interactions, for example a spin-orbital interaction, are also important because can mix different spin states inside single ion.

We consider spin crossover far from the thermodynamic phase transition in the paramagnetic phase, it allows us to simplify this interaction and substitute the effect of exchange with the effective mean field. This mean field is spatially uniform for the ferromagnetic insulator or two-sublattice for the antiferromagnet one. Examples given above  \cite{STL,GKL,GSL,DHO} correspond to the anti- or ferrimagnetics. In any case this mean field just renormalizes the interionic multielectron energies $E_1$ and $E_2$, and is irrelevant to the crossover phenomenon. Another interaction that is smaller then the exchange one is given by relativistic anisotropy contribution to the $H_{ij}$. For example a spin-orbital interaction can mix different spin states inside single ion, and it occurs to be important in our problem.

In what follows we will consider the simplified spatially uniform model \cite{NO1}. Motivation for this simplification is as follows. Despite that spin crossover is related to many body system, the essential features of its dynamics can be described by effective, Landau-Zener type, Hamiltonian \cite{DZ1}.

The Hamiltonian of our model is given by $ {\cal H} =\sum^N_{i=1} {\cal H}_i$, where
\begin{align}\label{eqH2a}
{\cal H}_i= \left(
     \begin{array}{cc}
       \lambda_0 & 0 \\
       0 & \lambda_0 \\
     \end{array}
   \right)
   + \left(
       \begin{array}{cc}
         \varepsilon &  \rho  e^{-i\varphi}\\
         \rho  e^{i\varphi} & -\varepsilon
       \end{array}
     \right).
  \end{align}
The energy spectrum is given by $\varepsilon_{\pm}= \lambda_0 \pm \sqrt{\varepsilon^2 + \rho^2}$.
Both $\lambda_0$ and $\varepsilon$ are pressure dependent. Further we assume that the spin excitation gap is given by
\begin{align}\label{eq1a}	
    \varepsilon(P) = \varepsilon_0\bigg(1 -\frac{P}{P_c}\bigg).
\end{align}
The crossover takes place at the point $P_c$ when $\varepsilon(P_c) = 0$. The spin-orbit coupling $\lambda= \rho  e^{i\varphi}$ (with $\rho \ll \varepsilon_0$) mixes the different spin states, and it plays the role of quantum fluctuations in our Ising pseudospin basis. The dimensionless spin gap, $P/P_c -1$, plays a role of relative temperature $(T/T_c - 1)$ near the critical point $T_c$ \cite{ZDZ}.

\section{ Quench dynamics induced by high pressure}

We consider time dependent Schr\"odinger equation for the Hamiltonian (\ref{eqH2a}) assuming for simplicity that the pressure is the linear function of time, $P= P_c(1 +  t/\tau_Q)$. Inserting this expression into Eq. (\ref{eq1a}), we obtain $\varepsilon (t)=  -\varepsilon_0 t/\tau_Q$. The parameter $\tau_Q$ depends on $\dot P$ and can be written as $\tau_Q= P_c/\dot P$.

Let $|1\rangle$ and $|0\rangle$ are eigenstates of the operator $\hat\sigma_z$, so that  $\hat\sigma_z |1\rangle = |1\rangle$ and $\hat\sigma_z |0\rangle =- |0\rangle$. Expressing a generic state vector as
\begin{align}\label{S1}
|\psi(t)\rangle =e^{-i\int\lambda_0(t)dt} ( C_1(t) e^{-i\varphi/2}|1\rangle 
+ C_0(t) e^{i\varphi/2}|0\rangle ),
\end{align}
we find that the coefficients $C_1(t)$ and  $C_0(t)$ satisfy the Schr\"odinger equation with the time-dependent Hamiltonian in the Landau-Zener (LZ) form (in units $\hbar = 1$)
\begin{align} \label{LZ}
i\frac{d}{d t}\left(
                  \begin{array}{c}
                    C_1(t) \\
                   C_0(t) \\
                  \end{array}
                \right)
=\left(
   \begin{array}{cc}
     -\Delta t& \rho \\
     \rho & \Delta t \\
   \end{array}
 \right)
\left(
\begin{array}{c}
                    C_1(t) \\
                   C_0(t) \\
                  \end{array}
                \right ),
\end{align}
where $\Delta = \varepsilon_0/\tau_Q$.

In terms of dimensionless scaled time $\tau = \sqrt{\Delta}t=\tau_0(P/P_c -1)$ with $\tau_0 = \sqrt{\varepsilon_0 \tau_Q}=\sqrt{\varepsilon_0 P_c/\dot P}$, the Landau-Zener model is described by the Hamiltonian
\begin{align}\label{H2}
\mathcal H =    \left(
 \begin{array}{cc}
    -\tau & \omega \\
    \omega & \tau\\
  \end{array}
\right),
\end{align}
where $\omega = \rho/\sqrt{\Delta} =\tau_0\rho/\varepsilon_0$ is the dimensionless coupling constant.
Writing $|u(\tau)\rangle =  C_1(\tau) |1\rangle +C_0(\tau) |0\rangle$, one can recast the Schr\"odinger equation (\ref{LZ}) as
\begin{align}\label{eq1}
  i\frac{d}{d\tau} |u(\tau)\rangle = {\mathcal H}(\tau)|u(\tau)\rangle.
\end{align}
Here the time $\tau$ runs from the initial time 
$\tau_i= -\sqrt{\varepsilon_0 \tau_Q}$, corresponding to the initial pressure $P_i=0$, to final $\tau_f=(P_f/P_c -1) \sqrt{\varepsilon_0 \tau_Q}$, corresponding to $P_f$ at the end of quench ($P_f > P_c$). Further we assume  that $\varepsilon_0 \tau_Q \gg 1$, then time $\tau$ can be extended to $\pm \infty$, and the problem becomes fully equivalent to the LZ problem.

The energy spectrum  of the Hamiltonian (\ref {H2}) is given by $\varepsilon_{\pm}(\tau)= \pm \sqrt{ \tau^2 + \omega^2 }$, and its instantaneous eigenvectors  can be written as
\begin{align} 
|u_{-}(\tau)\rangle = \left(\begin{array}{c}
-\sin\frac{\theta(\tau)}{2}\\
\cos \frac{\theta(\tau)}{2} \end{array} \right ), \,
|u_{+}(\tau)\rangle = \left(\begin{array}{c}
                  \cos\frac{\theta(\tau)}{2}\\
                  \sin\frac{\theta(\tau)}{2}
                  \end{array}\right) \label{r}
\end{align}
where $\cos\theta(\tau) = -\tau/\sqrt{\tau^2 + \omega^2} $. The energy gap between the ground and excited states equals $2\sqrt{\tau^2 + \omega^2}$. 

From Eq. (\ref{r}) it follows that while the ground state behaves at $\tau =\pm \infty$ as  $|u_{-}(-\infty)\rangle \rightarrow |0\rangle$ and $|u_{-}(+\infty)\rangle \rightarrow |1\rangle$, the excited state behaves as follows: $|u_{+}(-\infty)\rangle \rightarrow |1\rangle$ and $|u_{+}(+\infty)\rangle \rightarrow |0\rangle$. The state $|u_{-}(-\infty)\rangle$ corresponds to the high spin (HS) of the system and $|u_{+}(-\infty)\rangle$ corresponds to the low spin (LS). Thus, if the system initially was in the HS state, at the end of the evolution its ground state corresponds to the LS. Tunnelling between the positive and negative energy eigenstates, leading to the mixture of HS and LS, happens in the neighbourhood of the critical point $\tau_c =0$ ($P=P_c$) when $\tau \in (-\omega, \omega)$ \cite{DB}.

We assume further that the evolution of the system starts at the moment of time $\tau_i = -\sqrt{\varepsilon_0 \tau_Q}$ ($P(\tau_i)=0$) from the ground state $|u_{-}(\tau_i)\rangle$. Since $\rho \ll \varepsilon_0$, we have $|u_{-}(\tau_i)\rangle \propto |0\rangle$. This yields the following initial conditions: $C_0(\tau_i)=1$ and $C_1(\tau_i)=0$. At the end of evolution we obtain, $|u_{-}(\tau) \rightarrow|1\rangle$, while $\tau \rightarrow +\infty$. 

\begin{figure}[tbp]
\scalebox{0.4}{\includegraphics{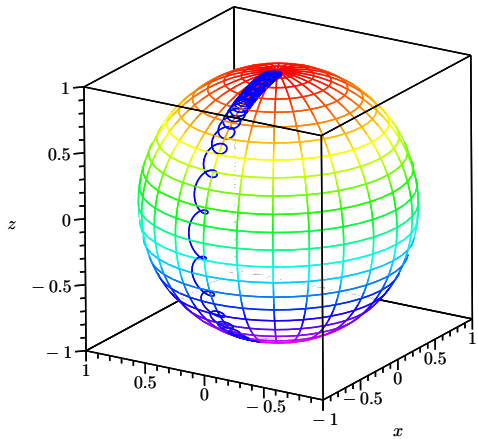}}
\caption{(Color online) Bloch vector's evolution ($\omega =3$). The motion starts at the south pole of the sphere and asymptotically ends at the north pole.}
\label{E1}
\end{figure}

\begin{figure}[tbp]
\scalebox{0.425}{\includegraphics{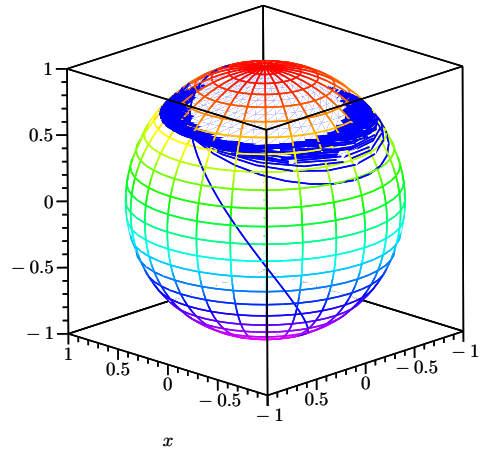}}
\caption{(Color online) Bloch vector's dynamics ($\omega =0.75$). The evolution starts at the south pole of the sphere.}
\label{E1a}
\end{figure}

\begin{figure}[tbp]
\scalebox{0.35}{\includegraphics{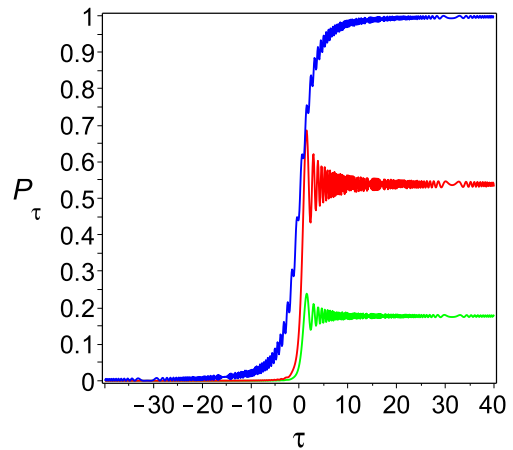}}
\caption{(Color online) Probability of transition $P_\tau$ obtained from exact solution of LZ problem as function of $\tau$. From up to down: $\omega= 3, 0.5, 0.25$.}
\label{P2a}
\end{figure}

In Figs. \ref{E1} -- \ref{P2a} we present the results of numerical solution of the Schr\"odinger equation. In Fig. \ref{E1}, \ref{E1a} time evolution of the Bloch vector, $\mathbf n= \langle u|\boldsymbol\sigma |u \rangle$, is shown.
The motion begins at the south pole of the two-dimensional sphere $S^2$ and for $\omega = 3$ ends at the north pole. However for the choice of the parameter $\omega < 1$, numerical simulation shows that the Bloch vector never reaches the  north pole, which corresponds to the LS state. This implies that at the end of evolution the quantum system does not remains in the ground state and its final state is the mixture of the LH and LS states. 

In Fig. \ref{P2a} the probability $P_\tau= |C_1(\tau)|^2$ of transition $|0\rangle$ $\rightarrow$ $|1\rangle $ is depicted for various values of $\omega$. As can be seen, with decreasing of adiabadicity parameter $\omega$  the transition probability decreases as well. Its asymptotic behaviour is described by the LZ formula (\ref{P}).

\subsection{Exact solution of the Landau-Zener problem}

The exact solution of the Eq.(\ref{eq1}) is given in terms of the parabolic cylinder functions\cite{abr,Leb,VGBM}, $D_{-1 -i\omega^2/2}(z)$, where $z= \sqrt{2}\,\tau e^{-i\pi/4}$ . Assuming that initially the system was in the ground state, $|u_{-}(-\infty)\rangle \rightarrow |0\rangle$, we obtain the initial conditions as follows: $C_0(-\infty)=1$ and  $C_1(-\infty)=0$. The probability of transition to the state $|1\rangle$ at time $\tau$ is given by \cite{VGBM,VN1,SSMO}
\begin{align}\label{eq2}
P_\tau = \frac{\omega^2}{2}e^{-\pi\omega^2/4}\big|D_{-1 - i\omega^2/2}(\tau \sqrt{2}e^{3i\pi/4})|^2.
\end{align}

As can be shown, the condition $\varepsilon_0 \gg \rho$ may be recast to $|\tau_i| \gg \omega$, where $\tau_i= -\sqrt{\varepsilon_0 \tau_Q}$ is the initial time. Since $|\tau_i| \gg 1$, for $\tau > |\tau_i|$ the so-called
{\em weak-coupling asymptotic} approximation may be applied \cite{VGBM}. The asymptotic of the transition probability is 
\begin{align}\label{P2}
P_\tau \sim 1- e^{-\pi\omega^2} - \frac{2\omega}{\tau}e^{-\pi\omega^2/2}\sqrt{1 -e^{-\pi\omega^2}} \cos\xi_w(\tau),
\end{align}
where
\begin{align}\label{xi}
  \cos\xi_w(\tau)= \frac{\pi}{4} +\tau^2  +\frac{\omega^2}{2}{\tau^2}\ln2 +\arg{\Gamma\bigg(1 -i\frac{\omega^2}{2}\bigg)}.
\end{align}

\subsubsection{Adiabatic approximation}

\begin{figure}[tbp]
\scalebox{0.35}{\includegraphics{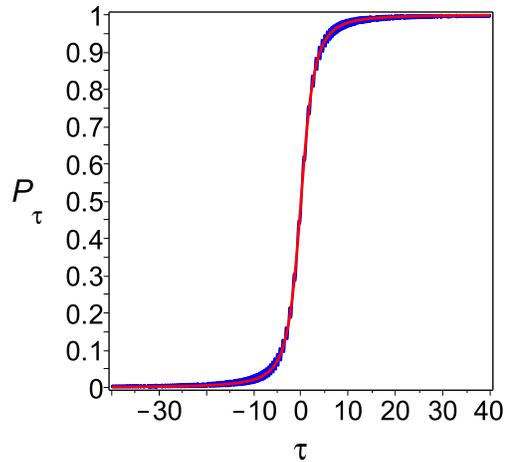}}
\caption{(Color online) Probability of transition $|0\rangle$ $\rightarrow$ $|1\rangle $ as function of dimensionless time $\tau$ ($\omega= 3$). Red line: probability of adiabatic transition, $P_{ad}(\tau)$. Blue line: probability of transition $P_\tau$ obtained from the exact solution of the LZ problem.}
\label{P2c}
\end{figure}

\begin{figure}[tbp]
\scalebox{0.35}{\includegraphics{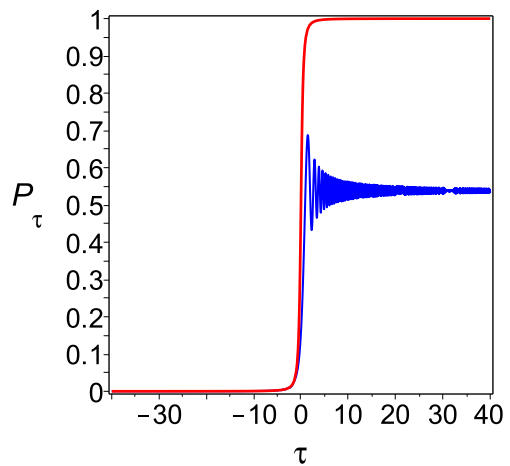}}
\caption{(Color online) Probability of transition $|0\rangle$ $\rightarrow$ $|1\rangle $ as function of $\tau$ ($\omega= 0.5$). Red (upper) line: probability of adiabatic transition. Blue (lower) line: transition probability obtained from exact solution of the LZ problem.}
\label{P2e}
\end{figure}

In the adiabatic approximation the probability of the system to remain in the ground state may be described by 
\begin{align}\label{P2b}
    P_{ad}(\tau)= |\langle u_{-}(\tau)|0\rangle|^2 = \frac{1}{2}\bigg(1 +\frac{\tau}{\sqrt{\tau^2 + \omega^2}}\bigg).
\end{align}
The condition for adiabatic evolution, required by the adiabatic theorem, is $\omega^2 \gg 1$ \cite{VN1}.
The probability of remaining in the ground state at the end of evolution ($\tau \rightarrow \infty$) is given by $P_{ad}= |C_1(+\infty)|^2$. For slow evolution we can use the LZ formula \cite{LL,ZC} to describe the probability of adiabatic evolution:
\begin{align}\label{P}
P_{ad} =1-e^{-\pi\omega^2}.
\end{align}
In Figs. \ref{P2c}, \ref{P2e} the probability of the adiabatic transition (red line) and the results of the exact solutions (blue line) are depìcted. As can be seen for $\omega =3$ there is a good agreement between exact solution and the adiabatic formula (\ref{P2b}). For $\omega=0.5$ the disagreement takes place.

Using the so-called {\em adiabatic-impulse} (AI) approximation \cite{DB,DZ1}, qualitatively, the dynamics of the Landau-Zener model can be described by the  Kibble-Zurek theory of nonequilibrium phase transitions \cite{KTW,ZHW,ZHW1}.
The AI-approximation assumes that the whole evolution can be divided in three parts and up to the phase factor the wave $|u(t)\rangle$ function approximately can be described as
\begin{align*}
   \tau \in [-\infty, - \hat \tau ] :  & & |u(\tau )\rangle\approx |u_{-}(\tau )\rangle  \\
   \tau  \in [ - \hat \tau,\hat \tau ] : & & |u(\tau )\rangle\approx |u_{-}(- \hat \tau )\rangle   \\
    \tau  \in [  \hat \tau , +\infty] :  & & |\langle u(\tau )|u_{-}(\tau )\rangle|^2 = \rm const
\end{align*}
where the time $\hat \tau $, introduced by Zurek \cite{ZHW}, is called the {\em freeze-out time} and define the instant when behaviour of  the system changes from the adiabatic regime to an impulse one where its state is effectively frozen and then back from the impulse regime to the adiabatic one.

If the evolution starts at moment $\tau_i \ll -\hat \tau$ from the ground state, the equation for determining $\hat \tau$ reads $\pi\hat \tau/2 = 1/\rm gap(\hat \tau)$ (for details of calculation see Ref. \cite{DZ1}), and its solution is given by
\begin{align}\label{LZ1}
\hat \tau = \frac{\omega}{\sqrt{2}} \sqrt{\sqrt{1+ \frac{4}{\pi^2\omega^4}}-1}.
\end{align}

Using the relation $\tau =\tau_0(P/P_c -1)$, we find that the change of the adiabatic regime to a non-adiabatic one occurs when the pressure is $P_1 = P_c\big(1 - \hat \tau /\tau_0\big)$, and the non-adiabatic evolution becomes the adiabatic evolution again when the pressure increases up to $P_2 = P_c\big(1 +  \hat \tau /\tau_0\big)$. From here we find that within the interval of pressure, $\Delta \hat P = P_2 - P_1 = 2P_c \hat \tau /\tau_0 $, the behaviour of the system is described by the impulse regime. 

Employing Eq. (\ref{LZ1}), we approximate $\Delta \hat P$ for fast ($\omega^2\ll 1$) and slow ($\omega^2\gg 1$) transitions as
\begin{align} \label{Eq6a}
\frac{\Delta \hat P}{P_c} = \left \{
\begin{array}{l}
\displaystyle \frac{1}{\sqrt{\pi}\tau_0}, \quad \omega^2 \ll 1 \\
\displaystyle \frac{1}{{\pi} \omega \tau_0}, \quad \omega^2 \gg 1
\end{array}
\right .
\end{align}

In the AI approximation the probability, $P_e$, of finding the system in the excited state at $\tau_f \gg \hat \tau$ can be calculated as follows \cite{DB,DZ1}:
\begin{align}\label{P3}
P_e \approx P_{AI} = |\langle u_{+}(\hat \tau)|u_{-}(-\hat\tau)\rangle|^2 = \frac{\hat\tau^2}{\omega^2 +\hat\tau^2}
\end{align}
Substituting $\hat \tau$ from (\ref{LZ1}), we obtain
\begin{align}\label{P3a}
    P_{AI} = \frac{2}{x^2 +x\sqrt{x^2 +4} + 2},
\end{align}
where $x= \pi\omega^2$. 

\begin{figure}[tbp]
\scalebox{0.35}{\includegraphics{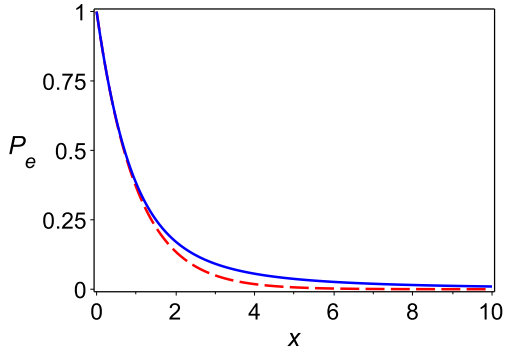}}
\caption{(Color online) Probability, $P_e$, of finding the system in the excited state ($x=\pi\omega^2$): blue line -- $P_e \approx P_{AI}$ (\ref{P3}), dashed red line -- the LZ expression, $P_e = e ^{-\pi \omega^2}$.}
\label{PAI}
\end{figure}

For $\omega^2\ll 1$, from Eq. (\ref{P3a}) it follows $P_{AI} \approx 1 - \pi \omega^2$. In the first order this coincides with the result predicted by exact LZ formula: $P_e = e ^{-\pi \omega^2}$. For the adiabatic evolution, $\omega^2 \gg 1$, we obtain $P_{AI} \approx 1/ \pi^2 \omega^4$ (See Fig. \ref{PAI}.). As can be seen, the AI approximation is good enough for $\omega^2 \leq 1$ and in the limit $\omega^2 \gg 1$.

{\em Comparison with experimental data. } -- To compare our theoretical finding with the experiments on spin crossover under the high pressure we use the data from Refs. \cite{STL,GKL,GSL,DHO}. The typical value of the critical pressure is $P_c = 50 \rm GPa$, and the rest of parameters are taken as follows:  $\varepsilon_0=  \, eV$, $\rho = 0.01 \,\rm eV$ and $\tau_Q \approx 10^4 \, \rm s$.  The computation yields: $\tau_0 \approx 10^9$ and $\omega \approx 10^7$. Thus the spin crossover under slowly changed pressure realized in the cited experiments is a highly adiabatic process ($\omega^2 \gg 1$).
Using (\ref{Eq6a}), we find the that the domain of non-adiabaticity is defined by $\Delta \hat P \approx 10^{-7} \,\rm Pa$. The corresponding interval of time  $\hat t \approx 1/(\pi \rho \tau_0)$ is  $\hat t \approx 10^{-21} \, \rm s$.

\subsection{Quench dynamics under shock-wave load}

In this section we study the quench dynamics in the spin system under time-dependent pressure, $P(t)$. We assume that at the initial moment of time $P(t_i)=0$, and at the end of evolution $P(t_f)=P_0$. Further it is convenient to present $P(t)$ as $P(t)= P_0 s(t)$, where $s(t_i)=0$ and $s(t_f)=1$. 

One can observe that the first term in the Hamiltonian (\ref{eqH2a}), yielding contribution to the total phase factor of the the wave function, does not affects the dynamics of the system and may be omitted. Setting for simplicity $\varphi=0$ and using Eq. (\ref{eq1a}), one can recast the time-dependent driving Hamiltonian as follows:
\begin{align}\label{eqH2c}
{\cal H}_\tau(t)= \left(
       \begin{array}{cc}
         \varepsilon_0(1- as(t)) &  \rho  \\
         \rho & -\varepsilon_0(1-as(t))
       \end{array}
     \right),
  \end{align}
where $a = P_0/P_c$. For a given $a$, the crossover occurs at the critical point $s_c = s(t_c)$ defined as $s_c = 1/a$.

In what follows we specify the pressure as a pulse with the shape determined by 
\begin{align} \label{Eq6}
P= \left\{
\begin{array}{l}
0, \quad t< 0 \\
P_0\tanh(\alpha t), \quad t \geq 0,
\end{array}
\right.
\end{align}
where $P_0$ is the pulse height. For $P_0 =P_c$, expanding $P(t)$ near of the critical point up to the first order, we obtain the related LZ problem:
 $P(t) = P_c(1+ \alpha t)$.
From here we obtain the corresponding LZ adiabaticity parameter as $\omega_{LZ} =  \rho/\sqrt{\alpha\varepsilon_0}$.

Applying the adiabatic theorem we find that condition
for adiabatic evolution can be written as follows:
\begin{align}\label{Eq1}
\frac{1}{\rho^2}\max_{0<P<P_0}\left |\langle u_+|\frac{d {\cal H}_\tau}{dt}|u_-\rangle \right | \ll 1.
\end{align}
Next using Eq. (\ref{Eq6}), we obtain
\begin{align}\label{Eq2}
\frac{1}{\omega_{LZ}^2}\max_{0<s<1}\left |\frac{a\beta (1-s^2)}{\sqrt{(1-as)^2 +\beta^2}}\right |\ll 1,
\end{align}
where $\beta=\rho/\varepsilon_0$. In terms of the function $\omega(s)$ defined as
\begin{align}
\frac{1}{\omega^2}= \frac{1}{\omega_{LZ}^2}\frac{a\beta (1-s^2)}{\sqrt{(1-as)^2 +\beta^2}},
\end{align}
the condition of adiabaticity (\ref{Eq2}) can be recast to the following:
\begin{eqnarray}
{\max_{0<s<1}}\bigg( \frac{1}{\omega^2(s)}\bigg) \ll 1
\end{eqnarray}

\begin{figure}[tbp]
\scalebox{0.3}{\includegraphics{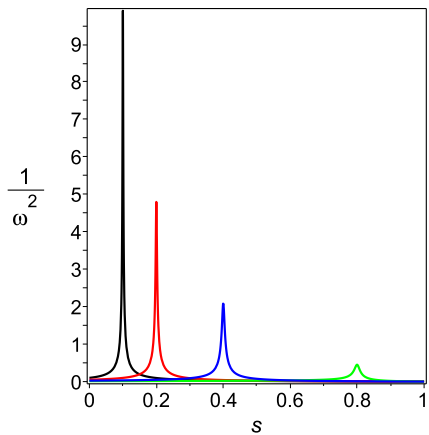}}
\caption{(Color online) Parameter of adiabaticity as function of $s$. From the right to the left: $a=10,5,2.5,1.25$ ($\omega_{LZ}= 1$).}
\label{P5}
\end{figure}
In Fig. \ref{P5} the function $1/\omega^2(s)$ is depicted ($\omega_{LZ}$ =1). We observe that $1/\omega^2$ has a maximum in the critical point. This is in agreement with the general observation on breaking-down of adiabaticity in the neighbourhood of the critical point.

\begin{figure}[tbp]
\scalebox{0.3}{\includegraphics{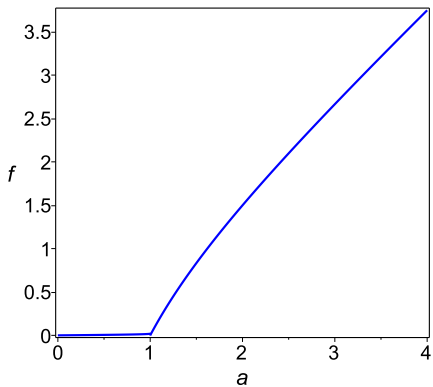}}
\caption{(Color online) Dependence of $f(a)$ on the amplitude $a=P_0/P_c$ ($\beta= 0.01$).}
\label{P5a}
\end{figure}

For $\beta \ll 1$ we obtain the following estimate
\begin{align}
\frac{1}{\omega^2} \approx\frac{f(a)}{\omega_{LZ}^2} ,
\end{align}
where
\begin{align}\label{Eq3}
f(a) = \left\{
\begin{array}{l}
\displaystyle\frac{2a \beta\sqrt{1-a^2}}{(1+\sqrt{1-a^2})\sqrt{1-a^2 +\beta^2}}, \quad a\leq 1 \\
 a- {1}/{a}, \quad a\geq 1
\end{array}
\right.
\end{align}
In Fig. \ref{P5a} the function $f(a)$ is depicted. We observe that $f(a) \ll 1$  for $a\leq 1$ and it is increasing function, $f(a) >1$ for $a>1$. 

\begin{figure}[tbp]
\scalebox{0.35}{\includegraphics{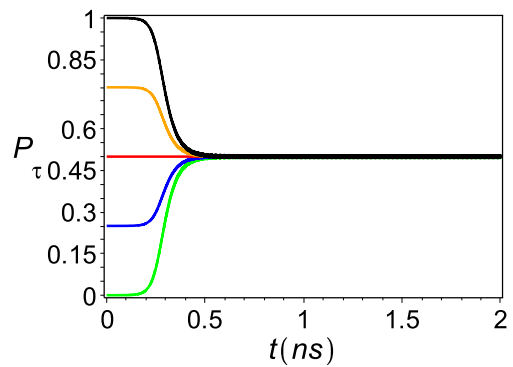}}
\caption{(Color online) Dependence of transition probability $P_\tau$ on time $t$, for different initial conditions ($P_0=P_c$). From up to down: $|C_0(0)|^2 = 0,0.25,0.5,0.75,1$ ($\alpha=10 \,\rm  ns^{-1}$).}
\label{P4c}
\end{figure}

\begin{figure}[tbp]
\scalebox{0.35}{\includegraphics{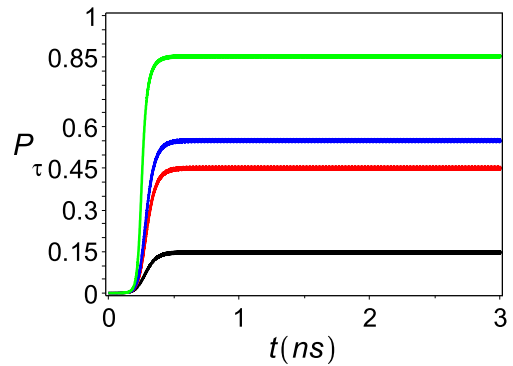}}
\caption{(Color online) Transition probability $P_\tau$ for different values of the pulse height as function of time $t$. From up to down: $a=P_0/P_c = 1.01,1.001, 0.999, 0.99$ ($\alpha=10 \,\rm ns^{-1}$).}
\label{P4f}
\end{figure}

\begin{figure}[tbp]
\scalebox{0.35}{\includegraphics{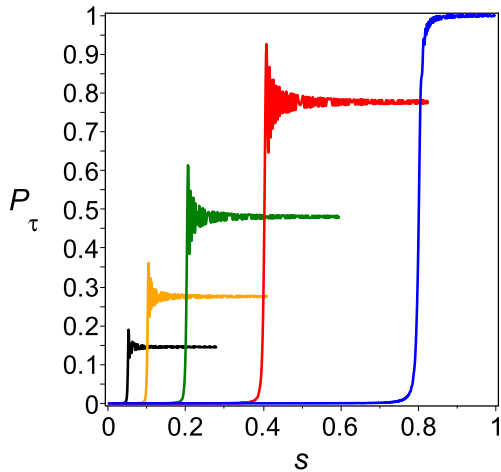}}
\caption{(Color online) Transition probability $P_\tau$ as function of scaled pressure $s$ for different values of the pulse height. From up to down: $a=P_0/P_c = 1.25, 2.5, 5, 10,20$ ($\alpha=10 \,\rm  ns^{-1}$).}
\label{P4e}
\end{figure}

In Figs. \ref{P4c} - \ref{P4e} we present the results of numerical calculations for different choice of parameter $a$. In all calculations the parameters were chosen as follows: $\varepsilon_0= 1\rm eV$, $\rho = 0.01 \rm eV$ and $\alpha=10 \, \rm ns^{-1}$. This yields $\omega_{LZ}=1$, and for $\beta=\rho/\varepsilon_0$, we obtain $\beta=0.01$.

Fig. \ref{P4c} shows that for $P_0 =P_c$, independently of the initial conditions, the final state of the system is defined by the equal mixture of LH and LS states. The reason for this phenomena is in growing quantum fluctuations which lead to the mixture of LH and LS states at the critical pressure $P_c$. 
In Figs. \ref{P4f} and \ref{P4e} the dependence of transition probability on time and pressure, respectively, is depicted for different values of the dimensionless amplitude of the pulse, $a= P_0/P_c$. In the interval of pressure, $0< P \lesssim 1.25 P_c$, the evolution is highly adiabatic, and for $P_0 \approx 1.25 P_c$ the system comes to the LS state with the minimal density of defects. However with the increasing of pressure amplitude, $P_0$, the probability to pass to the LS state is decreasing, and the system remains in the mixture of the HS ans LS states. Further increasing of the amplitude leads to the domination of the HS population over LS population.

\section{Summary and conclusions}

We have analytically studied the properties of spin crossover under the high pressure. We showed that at static loading and $P=P_c$ occupation numbers of both HS and LS states are equal at zero temperature ($T=0$), $n_{HS} = n_{LS} =0.5 $. For  $P<P_c$ we obtain $n_{HS} =1$ and $n_{LS } =0$, while for  $P>P_c$ one has $n_{HS} =0$ and $n_{LS } =1$. Static transition at $T=0$ is a sharp quantum phase transition with geometric Berry-like phase being the order parameter \cite{NO1}. Finite temperature removes singularity in the $n_{HS} (P)$ dependence. Thermal fluctuations between two states $|0 \rangle$ and $|1\rangle$  results in a smooth crossover instead of the quantum phase transition at $T=0$. 

To verify our theoretical predictions we have performed numerical simulations. Fig. \ref{P4f} shows that the temporal quantum fluctuations have an effect similar to the thermal fluctuations. Small deviations of the shock wave amplitude $P_0/P_c$ from the unity results not in a sharp change of the probability $P_\tau$ either to zero or to unity but to a continuous deviation of the $P_\tau$ from the 0.5 value. At the same time for $P_0/P_c =1$ any initial distribution of the HS and LS states will end its evolution in the equilibrium state $n_{HS} = n_{LS} =0.5 $ (Fig.\ref{P4c}). This conclusion is valid for any choice of parameter $\alpha$.

Another dynamical effect we would like to discuss is the adiabaticity violations near the critical pressure (Fig.\ref{P4e}). This is a manifestation of the general Kibble-Zurek theory. The results shown in Fig.\ref{P4e} are contra intuitive at the first glimpse. The shock wave with larger amplitude has smaller final probability for spin crossover and larger probability to stay in the initial HS state. To understand this effect one should note that the characteristic scale of the pressure increase is given by the factor $a\alpha$ for the wave with amplitude $a$. Thus larger amplitude wave also is faster.

\section*{Acknowledgments}

This research was supported by the President of Russia Grant NSh-1044.2012.2, Presidium of the Russian Academy of Science Project 2.16, RFBR  Grant 12--02-90410{\_}a, A.I. Nesterov
acknowledges the support from the CONACyT, Grant No. 118930.

\end{document}